\documentclass[a4paper]{spie}  

\usepackage{amsmath,amsfonts,amssymb}
\usepackage{graphicx}
\usepackage[colorlinks=true, allcolors=blue]{hyperref}
\usepackage[super]{nth}

\usepackage{lipsum}

\title{Tissue Cross-Section and Pen Marking Segmentation in Whole Slide Images}

\author[a,b]{Ruben T. Lucassen}
\author[a]{Willeke A.M. Blokx}
\author[b]{Mitko Veta}
\affil[a]{Department of Pathology, University Medical Center Utrecht, the Netherlands}
\affil[b]{Department of Biomedical Engineering, Eindhoven University of Technology, the Netherlands}

\authorinfo{Send correspondence to R.T.Lucassen. E-mail: r.t.lucassen@umcutrecht.nl}

\begin{document} 
\maketitle

\begin{abstract}
Tissue segmentation is a routine preprocessing step to reduce the computational cost of whole slide image (WSI) analysis by excluding background regions. Traditional image processing techniques are commonly used for tissue segmentation, but often require manual adjustments to parameter values for atypical cases, fail to exclude all slide and scanning artifacts from the background, and are unable to segment adipose tissue. Pen marking artifacts in particular can be a potential source of bias for subsequent analyses if not removed. In addition, several applications require the separation of individual cross-sections, which can be challenging due to tissue fragmentation and adjacent positioning. To address these problems, we develop a convolutional neural network for tissue and pen marking segmentation using a dataset of 200 H\&E stained WSIs. For separating tissue cross-sections, we propose a novel post-processing method based on clustering predicted centroid locations of the cross-sections in a 2D histogram. On an independent test set, the model achieved a mean Dice score of 0.981\,$\pm$\,0.033 for tissue segmentation and a mean Dice score of 0.912\,$\pm$\,0.090 for pen marking segmentation. The mean absolute difference between the number of annotated and separated cross-sections was 0.075\,$\pm$\,0.350. Our results demonstrate that the proposed model can accurately segment H\&E stained tissue cross-sections and pen markings in WSIs while being robust to many common slide and scanning artifacts. The model with trained model parameters and post-processing method are made publicly available as a Python package called \textit{SlideSegmenter}.
\end{abstract}


\section{INTRODUCTION}
\begin{figure}[t]
\centering
\includegraphics[width=\textwidth]{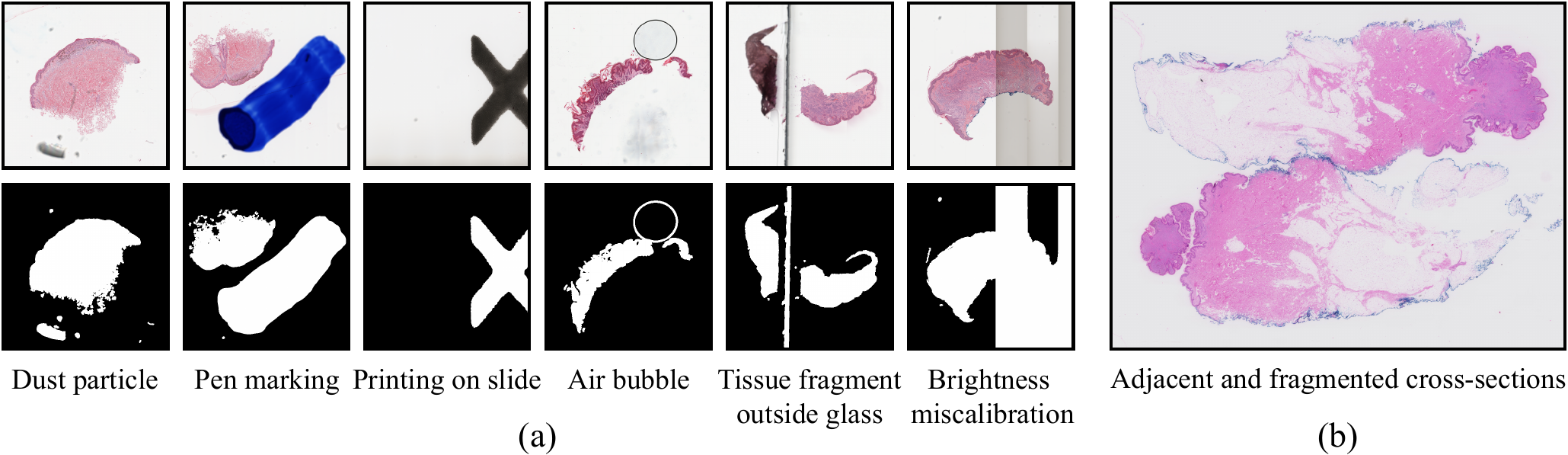}
\caption{(a) Examples of slide and scanning artifacts cropped from WSIs (top row) and corresponding segmentation results using \texttt{MorphologicalMasker} from the TIAToolbox\cite{pocock2022tiatoolbox} (version 1.5.1) with default parameter values (bottom row). (b) Example of two tissue cross-sections with adjacent positioning and fragmentation.}
\label{dataset}
\end{figure}

Tissue segmentation is a routine preprocessing step for most whole slide image (WSI) analysis tasks in computational pathology. By separating tissue from the background, which is generally uninformative, the computational cost of subsequent analysis can be reduced substantially. A common approach for tissue segmentation is the use of traditional image processing techniques such as thresholding followed by morphological operations \cite{lu2021data, pocock2022tiatoolbox}. Despite their simplicity and efficiency, these methods often require manual adjustments to parameter values for atypical cases, fail to exclude all slide and scanning artifacts from the background (see Figure \ref{dataset}a), and are unable to segment adipose tissue. With research moving towards the use of enormous datasets \cite{chen2023general, vorontsov2023virchow}
and with pathology departments starting to adopt computer-aided tools in their digital workflow \cite{van2023deep}, automated and robust tissue segmentation without the need for manual corrections is a necessity.

In traditional pathology workflows, some pathologists use pen markings on the glass slides to highlight regions of interest. If not removed before scanning, these markings also appear on the corresponding WSIs. Unlike other artifacts such as air bubbles and dust particles which appear in random locations during slide preparation and scanning, most pen markings are strongly correlated with nearby tissue regions of diagnostic importance. Because deep learning models are susceptible to learning spurious correlations from the training data \cite{winkler2019association}, using WSIs with pen markings for model development can introduce bias in downstream tasks like detection or classification of tumor tissue. Hence, the capability to identify and exclude pen marking artifacts from the tissue segmentation is particularly important. 

In dermatopathology, due to the relatively small size of many of the skin biopsies and excisions, it is common practice to position multiple tissue cross-sections on a single slide. The separation of cross-sections in WSIs can be challenging due to tissue fragmentation and adjacent positioning (see Figure \ref{dataset}b), yet is relevant for further analysis. For example, it enables the use of positional information from tissue regions within cross-sections, while neglecting the often arbitrary or inconsistent positioning between cross-sections. Moreover, it allows for image registration from one cross-section to another.

To this end, we develop a convolutional neural network (CNN) for tissue and pen marking segmentation in hematoxylin and eosin (H\&E) stained WSIs, and propose a novel post-processing method for separating tissue cross-sections when multiple are present on a single WSI. The model with trained model parameters and post-processing method are made publicly available as a Python package called \textit{SlideSegmenter}\footnote[1]{\url{https://github.com/RTLucassen/slidesegmenter}}.

\section{METHODS}
\subsection{Dataset}
The dataset used in this work consists of 200 WSIs of H\&E stained skin biopsies and excisions with melanocytic lesions, accessioned between January 1, 2013 and December 31, 2020, and obtained from the digital archive of the Department of Pathology at the University Medical Center Utrecht, the Netherlands. The dataset was in part randomly selected and further enriched for WSIs with artifacts and fragmented or adjacently positioned tissue cross-sections. The study was conducted in compliance with the hospital's research ethics committee guidelines. All WSIs were de-identified.

The dataset was randomly divided on a patient-level into sets of 140, 20, and 40 WSIs for training, validation, and testing, respectively. A total of 36 WSIs were from slides with pen markings present. All WSIs were acquired using either a ScanScope XT scanner (Aperio, Vista, CA, USA) in DICOM format at 20$\times$ magnification with a resolution of 0.50 \textmu m per pixel (slides scanned before 2016) or a Nanozoomer 2.0-XR scanner (Hamamatsu photonics, Hamamatsu, Shizuoka, Japan) in NDPI format at 40$\times$ magnification with a resolution of 0.22 \textmu m per pixel (slides scanned after 2016). We worked with the WSIs at a magnification of 1.25$\times$ for model development and evaluation, which was selected as a trade-off between image detail and computational cost. At this magnification, the images ranged in size between 0.1-17.3 megapixels. All pen markings not superimposed on tissue were cropped from the training data and saved as a separate set of images to be used during model training. We developed and used a Python-based tool for bitmap annotation\footnote[2]{\url{https://github.com/RTLucassen/annotation_tool}}. All images were annotated by a single annotator (R.L.) and reviewed by a pathologist (W.B.).

\begin{figure}[t]
\centering
\includegraphics[width=\textwidth]{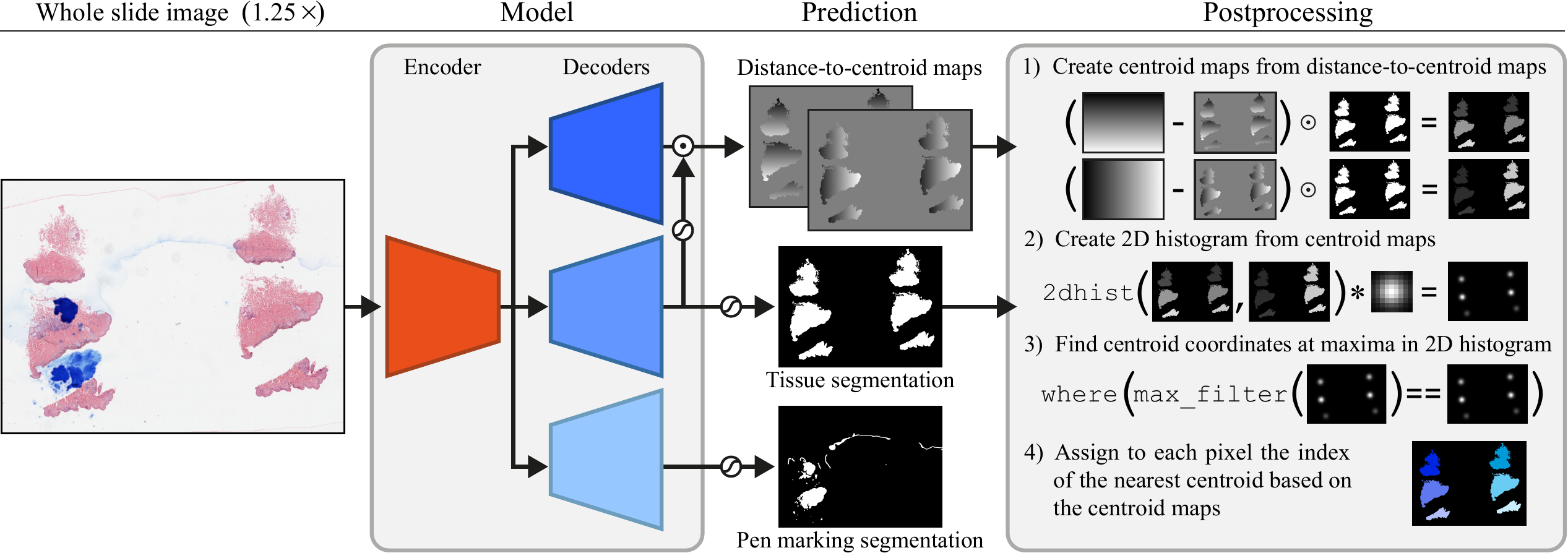}
\caption{Overview of the proposed method for tissue cross-section and pen marking segmentation.}
\label{overview}
\end{figure}

\subsection{Network Architecture and Post-Processing}
Our approach was inspired by HoVer-Net\cite{graham2019hover}, which is a method for cell nuclei segmentation and classification. HoVer-Net uses a CNN to predict for each pixel in the input image whether it belongs to a nucleus (i.e., binary segmentation), the relative horizontal and vertical distance with respect to the centroid of the nucleus, and the type of cell nucleus. As for post-processing, after constructing the energy landscape and instance marker images from the predicted segmentation and distance maps, a marker-controlled watershed algorithm is used to separate individual cell nuclei. This post-processing method was specifically designed to be capable of separating overlapping nuclei. However, when considering the task of separating segmented tissue cross-sections, not only can the segmentation of two adjacent cross-sections be connected, but a single cross-section can also be fragmented into multiple disjoined segments (see Figure \ref{dataset}b). Because each disjoined segment would be considered a separate instance in the watershed procedure, the post-processing by HoVer-Net does not translate effectively to our problem.

Hence, we developed a different post-processing algorithm, capable of both separating connected cross-sections and joining disconnected fragments of a single cross-section. An overview of the proposed method can be seen in Figure \ref{overview}. We first trained a CNN to predict for each pixel in the input image whether it is part of a tissue cross-section and/or pen marking, as well as the horizontal and vertical distance with respect to the centroid of the corresponding cross-section. Unlike the CNN in HoVer-Net, which predicts the relative distance with respect to the centroid in a range of -1 to 1, our model predicts the absolute horizontal and vertical distance in pixels. 
The post-processing algorithm can be divided into four steps, corresponding to the steps illustrated in Figure \ref{overview}: 
\begin{enumerate}
    \item The horizontal and vertical distance-to-centroid maps as predicted by the CNN are subtracted from the horizontal and vertical pixel coordinate maps, respectively, and multiplied element-wise with the tissue segmentation image to mask the background. In the resulting centroid maps, each foreground pixel value is equal to the predicted horizontal and vertical centroid location of the cross-section it is part of.
    \item The horizontal and vertical centroid maps, after excluding all background pixels, form the input to a 2D histogram, which shows the distributions of the predicted centroid locations. The height and width of the 2D histogram are both $k$~=~20$\times$ smaller than the height and width of the image. To prevent multiple maxima in the distribution for a single centroid location due to imperfect predictions, the 2D histogram is convolved with a Gaussian filter with standard deviation $\sigma$~=~2.
    \item Non-maximum suppression is performed on the 2D histogram. The maximum of each predicted centroid location distribution is determined by element-wise comparison of the 2D histogram before and after application of a maximum filter with a kernel of size $s$~=~15$\times$15\,px$^{2}$. The histogram bins that were identified as maxima must also exceed a threshold value $t$ equal to the \nth{98} percentile of the bin counts.
    \item The Euclidean distance is computed between all tissue pixels in the centroid map from step 1 and the centroid locations at the identified maxima in the 2D histogram from step 3. Each pixel is assigned to the cross-section with the nearest centroid.
\end{enumerate}

Given accurate distance-to-centroid predictions, all segments from one fragmented cross-section are mapped to the same 2D histogram location and are therefore no longer disjoined. Additionally, the pixels at the interface between two adjacent cross-sections are mapped to different locations in 2D histogram, generally making the cross-sections better separable. Hyperparameter values for the post-processing method were determined using a grid search based on the validation set after the CNN was trained.

The CNN architecture was based on U-Net\cite{ronneberger2015u} with several modifications. It consists of a single encoder attached to three decoders with shortcut connections. The encoder and decoders have five down- and upsampling layers, respectively. To increase the receptive field, which is required to accurately predict the distance to the centroid for large cross-sections, feature maps are down- and upsampled by a factor of 4 instead of 2 at all levels of spatial resolution except for the top level. The decoders for tissue and pen marking segmentation use a sigmoid as final activation function. The distance-to-centroid maps, predicted by the decoder without a final activation function, are multiplied element-wise with the predicted tissue segmentation map.

\subsection{Network Training}
The encoder and decoders of the CNN were jointly optimized based on the same loss functions as were used for training HoVer-Net. For the segmentation predictions, the network was trained to minimize the sum of the Dice loss and cross-entropy loss (weighted by a factor of 10). For the horizontal and vertical distance-to-centroid predictions, the network was trained to minimize both the mean squared error between the predicted and ground truth distances, as well as the mean squared error between the gradient of predicted and ground truth distances. Because of large differences in size between images in the dataset, which is also to be expected in a deployment environment, only a single image of variable size was used per batch. The network was trained starting from randomly initialized parameters for 100,000 iterations with gradients accumulated over every 5 iterations using the AdamW\cite{loshchilov2019decoupled} optimization algorithm ($\beta_1$~=~0.9, $\beta_2$~=~0.999). The learning rate was 3\,$\cdot$\,10\,\textsuperscript{-4} at the start and halved after every 20,000 iterations. The network parameters that resulted in the smallest loss on the validation set were saved, which was evaluated after every 500 iterations. 
Data augmentation was performed on the fly by randomly applying translations, rotations, scaling, flips, Gaussian noise, Gaussian blurring, JPEG compression, as well as shifts in the hue, saturation, brightness, and contrast. In addition, augmented versions of pen markings extracted from the training data were superimposed in random locations during training to increase the frequency and variation. Images were padded such that the height and width were equal to the nearest multiple of 512 pixels and randomly cropped to a maximum of 3072 pixels if necessary. Hyperparameters for model development were tuned based on the validation set results. The network and training procedure were implemented in the Pytorch framework\cite{paszke2019pytorch}.

\section{RESULTS}
\begin{figure}[t]
\centering
\includegraphics[width=\textwidth]{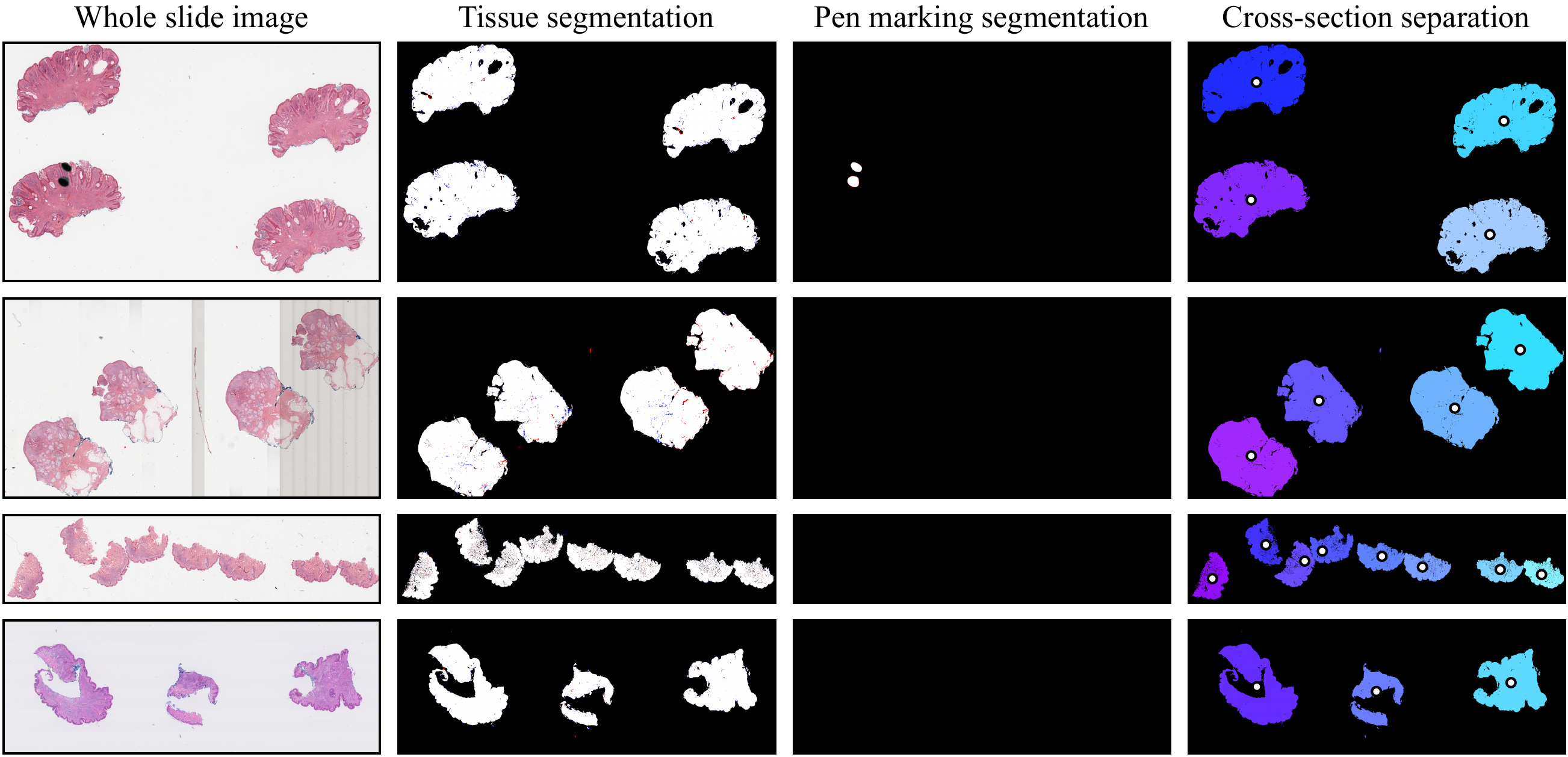}
\caption{Results for four WSIs from the test set. The segmentation results show true positives in white, true negatives in black, false positives in red, and false negatives in blue. The rightmost column shows the separated cross-sections indicated by different colors with the predicted centroid locations as white dots.}
\label{results}
\end{figure}

Results are reported as the mean and standard deviation based on all WSIs in the independent test set unless otherwise specified. The model achieved a Dice score of 0.981\,$\pm$\,0.033 for tissue segmentation. For pen marking segmentation, the Dice score was 0.912\,$\pm$\,0.090 based on 8 out of the 40 WSIs in the test set with pen markings. The model made false positive predictions in 8 out of the remaining 32 images absent of pen markings, of which 5 cases were considered minor ($\sim$10\,\textsuperscript{2} pixels whereas all true pen markings exceed 10\,\textsuperscript{4} pixels) and 3 cases were clear mistakes.
Regarding the separation of cross-sections, the model in combination with the post-processing method reached a mean absolute difference of 0.075\,$\pm$\,0.350 between the number of annotated and separated cross-sections. The segmentation and separation results for four WSIs from the test set are shown in Figure~\ref{results}. The model has for example learned to segment pen markings superimposed on tissue (first row), as well as to neglect scanning artifacts and to include adipose tissue regions (second row). Furthermore, based on the distance-to-centroid predictions by the model, the post-processing method is able to separate adjacent cross-sections (third row) and join fragmented cross-sections (fourth row).

\section{DISCUSSION AND CONCLUSION}
Our results demonstrate that the proposed model can accurately segment H\&E stained tissue cross-sections and pen markings in WSIs while being robust to many common slide and scanning artifacts. We believe that the segmentation model has the potential to replace its traditional counterpart in WSI preprocessing routines. Possible advantages of this would include a reduced need for manual corrections, as well as the prevention of bias by excluding pen markings and the separation of cross-sections for downstream tasks. These benefits are especially relevant considering the current trends of growing dataset sizes for research\cite{chen2023general, vorontsov2023virchow} and the starting adoption of computer-aided tools in the digital workflow of pathology departments \cite{van2023deep}.

Despite the strong segmentation performance as evidenced by the high Dice scores, there is still room for improvement. Visual inspection of the tissue segmentation results revealed that the remaining segmentation error can be attributed to false negative predictions for small regions of adipose tissue and false positive predictions for holes or tears in the tissue as well as for tissue debris at a distance from any cross-section. Moreover, artifacts in the background were on two occasions mistaken for tissue by the model. For pen marking segmentation, clear false positive predictions were identified in case of a large dust particle, a printed symbol on the slide, and a thick corneal layer of acral skin. Incorrect distance-to-centroid predictions were primarily observed in some of the largest cross-sections, which are underrepresented in the dataset and likely to be the most challenging due to the extensive receptive field required. Most of these errors can probably be explained by the still limited variation in the appearance of tissue and artifacts in the training data. We expect that the performance can be improved by increasing the training dataset with selected challenging cases and retraining the model.

Related methods for tissue and pen marking segmentation have primarily been developed for the purpose of quality control after scanning\cite{janowczyk2019histoqc, smit2021quality}, operating at a higher magnification and aiming to segment more types of artifacts, whereas our approach was designed to be used as part of preprocessing pipelines for subsequent analyses of WSIs. Currently, the model has only been trained on WSIs with H\&E stained tissue from skin biopsies and excisions. Exploratory analysis showed promising generalization to other tissue types as well, but this has not yet been evaluated quantitatively. Extending the model to segment tissue in WSIs with immunohistochemical stains, which might be even more challenging for traditional image processing techniques due to the smaller difference in intensity values between tissue and background, is a potential direction for future work. 

In conclusion, we developed a CNN for tissue and pen marking segmentation in H\&E stained WSIs, combined with a novel post-processing method for separating tissue cross-sections when multiple are present on a single WSI. The model with trained model parameters and post-processing method are made publicly available as a Python package called \textit{SlideSegmenter}.

\acknowledgments  
This research is financially supported by the Hanarth Fonds.

\bibliography{report} 
\bibliographystyle{spiebib} 

\end{document}